\documentclass[useAMS,usenatbib]{mn2e}

\usepackage{graphicx}
\usepackage{dcolumn}
\usepackage{bm}

\newcommand{\sss}{\scriptscriptstyle}
\newcommand {\be}{\begin{equation}} 
\newcommand{\ee}{\end{equation}}    

\def\vti{v_{{\sss T}i}}

\def\vte2{v_{{\sss T}\alpha}}

\title[Charge exchange in fluid description  of partially ionized plasmas]{Charge exchange in fluid description  of partially ionized plasmas}
\author[J. Vranjes, M. Kono,  and M. Luna]{J. Vranjes$^{1}$\thanks{E-mail:
jvranjes@yahoo.com}, M. Kono$^{2}$\thanks{E-mail:
kono@fps.chuo-u.ac.jp} and M. Luna$^{1}$\thanks{E-mail:
mluna@iac.es}
\\
$^{1}$Instituto de Astrof\'{\i}sica de Canarias, 38205 La Laguna, Tenerife, Spain and\\
Departamento de Astrof\'{\i}sica, Universidad de la Laguna, 38205 La Laguna, Tenerife, Spain \\
$^{2}$Faculty of Policy Studies, Chuo University, Tokyo, Japan}
\begin{document}

\date{Accepted 1988 December 15. Received 1988 December 14; in original form 1988 October 11}

\pagerange{\pageref{firstpage}--\pageref{lastpage}} \pubyear{2002}

\maketitle

\label{firstpage}

\begin{abstract}
 The effects of charge exchange on waves propagating in weakly ionized plasmas are discussed. It is shown that for low-frequency processes, ions and neutrals should be treated as a single fluid with some effective charge on all of them. We have derived a new momentum equation which should be used in such an environment. As a result, the low-frequency magnetic waves can propagate even if particles are not magnetized, which is entirely due to the charge exchange and the fact that it is not possible to separate particles into two different populations as charged and neutral species. So there can be no friction force between ions and neutrals in the usual sense. The mean force per particle is proportional to the ionization ratio $n_i/(n_i+ n_n)$.   Regarding the application of the theory to the  Alfv\'{e}n wave propagation in the lower solar atmosphere, the results predict that the plane of displacement of the fluid must change by 90 degrees when an Alfv\'{e}n wave propagates from the area where particles are un-magnetized (photosphere) to the area where they are magnetized (chromosphere). Because of the most accurate cross sections which we have here, it is possible to very accurately determine altitudes at which such rotation of the Alfv\'{e}n wave takes place.
\end{abstract}


\section{ Introduction}
Charge exchange (also known as charge transfer, or electron capture) is a process in which an ion and an atom exchange electron(s) in the process of collision. In noble gasses the charge exchange is more dominant than the usual elastic scattering \citep{raiz}. In hydrogen plasmas, like in the case of the lower solar atmosphere, the charge exchange takes place at a very high rate as well \citep{v1,v2}; see more about it further in the text.

Nevertheless, in the past the charge exchange has been completely ignored in the solar plasma literature. This in spite of the fact that the theory has been well known for more than 60 years \citep{dal, raiz, brand, eic}, and it has been thoroughly studied in the atomic physics and in the general plasma theory \citep{k1, k2, k3}. In fact, to the best of our knowledge, our two references given above are the only published works where the charge exchange has been studied in detail in the context of the solar plasma.

In what follows we are going to show that the charge exchange may be a crucial factor in explaining the propagation of waves in weakly ionized plasmas, in particular in the application to the lower solar atmosphere.

The reason for focusing onto the charge exchange is the following. Using standard  descriptions, in the lower solar atmosphere the ion friction with neutrals turns out to be  so strong that any magnetic perturbation is almost instantly destroyed \citep{v2, v3}. So an enormously strong magnetic field would be needed to have a magnetic wave propagating. The term magnetic waves is used here to describe the Alfv\'{e}n wave, and any other wave which requires magnetized particles.

Related to this, there is a problem of magnetization. As shown in \citet{v1}, the proton  collision frequency can exceed $10^9$ Hz. As such, it is several orders of magnitude above the proton gyrofrequency even if we assume strong magnetic fields of the order of 0.1 T. Under such conditions the motion of an individual particle is similar to brownian motion.  Though electrons can be magnetized and there may be a relative drift between the protons and electrons in the direction transverse to the magnetic field. This causes the Hall effect which consequently may affect protons as well although they remain unmagnetized, $\nu_i>\Omega_i$. This is in fact manifested as  the Hall term in Eq.~(\ref{d3})  later in the text.

It is therefore necessary to address the issue of magnetic waves in such a weakly ionized and highly collisional environment from a completely different perspective, by applying the charge exchange effects in a  self-consistent manner. In the present case this implies that particles change their identity so frequently that it becomes impossible to distinguish what is a charged particle and what is a neutral. In practical terms, they have a dual nature all the time and, in average, they all take part in electromagnetic perturbations. This turns out to be crucial, the friction effectively vanishes and electromagnetic perturbations can propagate. But there is a price for this: the effective electromagnetic force of a wave acts on all particles (ions and neutrals) at any time, so the total mean force felt by an individual particle can be drastically reduced.  All these effects are described in the text below.

\section{Establishing facts}

The conductivity in plasmas is usually dominated by electrons [$\sigma=e^2 n_e/(m_e \nu_e)$, where the notation is obvious, and $\nu_e$ is the electron total collision frequency] because of their much higher mobility as compared to ions. But the collisions in the solar photosphere are enormously frequent, and the electron collision frequency may be over $10^{10}$ Hz \citep{v1}. So the conductivity is very low, and it is in the range $1-30$ S/m. Compare this with the conductivity of the terrestrial sea water, which is around 5 S/m, i.e., similar to the photosphere. We know that ordinary fluid dynamics is perfectly able to describe phenomena in terrestrial oceans. So the question is why such a fluid theory is not good enough in the case of the solar photosphere? We shall come to this question again in the text below.

\begin{figure}
   \centering
    \includegraphics[height=5cm,bb=15 14 310 210,clip=]{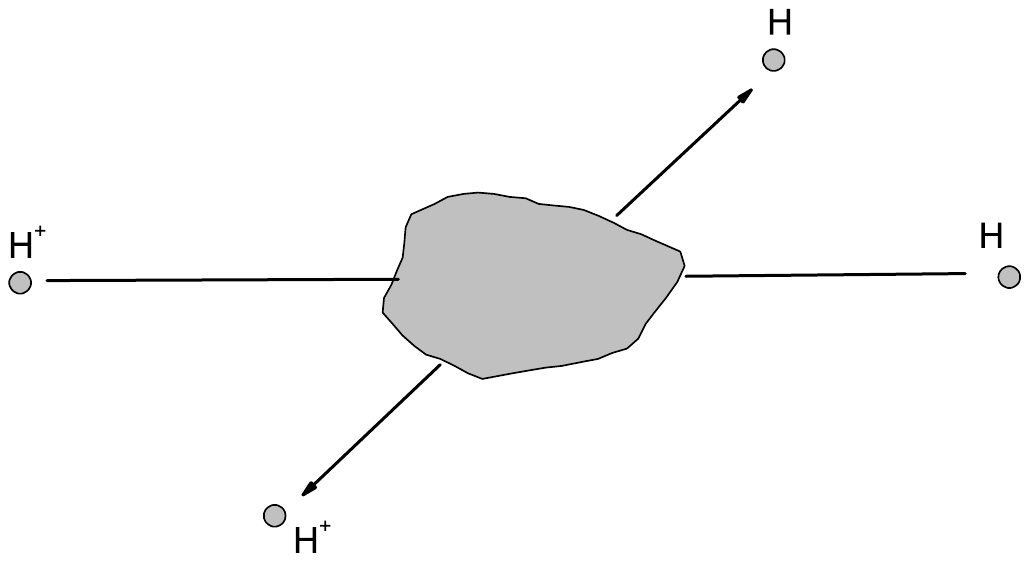}
   \caption{Which is which? Collision of a hydrogen atom and a proton in the presence of charge exchange, demonstrating the impossibility to distinguish particles exiting the collision zone. } \label{f1}
       \end{figure}

In the case of collisions of ions with their parental atoms, the quantum-mechanical effect of indistinguishability applies. A sketch of this effect is given in Fig.~\ref{f1}. When an ion and an atom collide, and we detect the particles after collisions, there  is no way to know which is which unless they are somehow additionally labeled. In other words, we may have the usual elastic scattering as well as the charge exchange. This is the meaning of the indistinguishability effect caused by the charge exchange, which therefore  must be consistently included in the theory \citep{k1, k2, k3}. This implies that the usual collision cross sections for elastic scattering, momentum transfer, and viscosity must include the charge exchange effect. Such cross sections are provided for the solar hydrogen plasma in our recent papers \citet{v1,v2}.

{\em Consequence 1:} The important conclusion which follows from this is that there will be {\em only one friction force term} in the ion momentum equation, instead of two (or even more) as can  frequently be seen in the plasma physics literature. The same holds of course for the momentum equation of neutrals. This all holds for relatively low energies, roughly speaking below 1 eV. Note that this energy range in fact covers most cases of interest, because at higher energies the amount of neutrals may be negligible and friction plays no important role.

\begin{figure}
   \centering
    \includegraphics[height=6.5cm,bb=15 14 272 224,clip=]{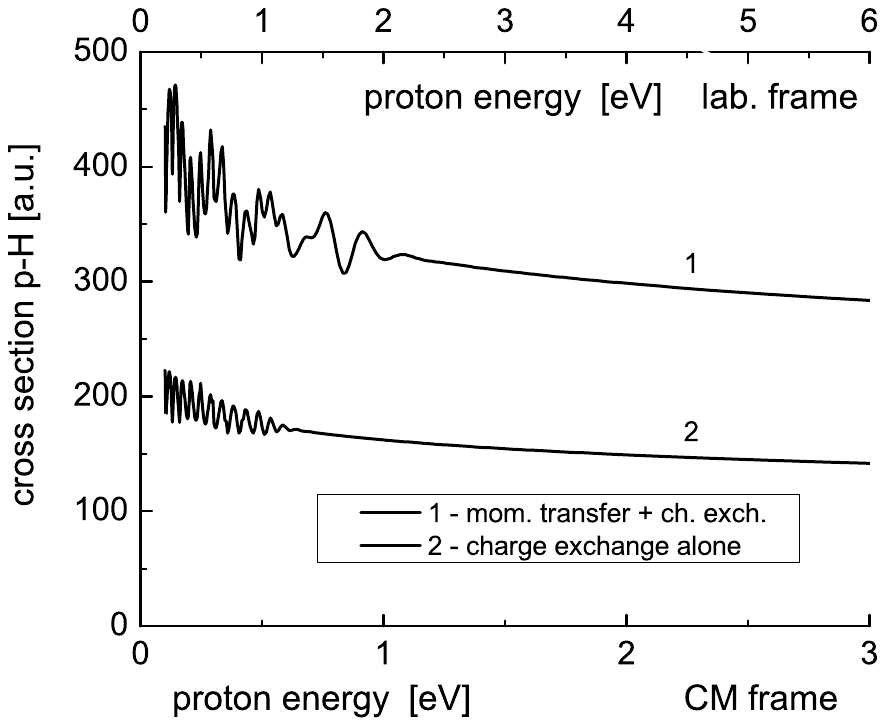}
   \caption{Cross sections for charge transfer (line 2), and for the momentum transfer with the effects of charge transfer included (line 1). Atomic unit is the square of the Bohr radius, a.u$=r_b^2= 2.8\cdot 10^{-21}$ m$^2$. Data adapted from \citet{k2, v1}. } \label{f2}
       \end{figure}

 Let us now see more quantitatively how important the change exchange is. We shall first discuss the cross section for this process. A lot of details are available in papers by Krstic and his group \citep{k2,k3}, and those have been used in our papers as well \citep{v1,v2}.  In Fig.~\ref{f2} we give the momentum transfer cross section affected by the charge transfer (line 1), and the charge transfer alone (line 2). The total correct value for the momentum transfer is increased due to the charge exchange roughly by a factor 2. Note that the charge transfer affects the other cross sections as well, i.e., those for elastic scattering and viscosity \citep{v1}. The variations of the lines in Fig.~\ref{f2} at low energy is the usual quantum mechanical effect.

  In the case of the solar photosphere and chromosphere, in many situations it may be appropriate to use the approximate value for the momentum transfer $  \sigma_{mt}\approx 200$ a.u., where a.u. is the square of the Bohr radius $\approx 2.8\cdot 10^{-21}$ m$^2$. Using a value which varies with the altitude (i.e., with the temperature) can be done  but for some problems [like the Alfv\'{e}n wave propagation studied by \citet{v3}] this will not have a profound effect on the results.

{\em Consequence 2:} Using the values for the momentum transfer presented by line 1 in Fig.~\ref{f2}, will most consistently include the effects of charge transfer in the friction force term in the momentum equation.

 Having so accurate cross sections we can now calculate the collision frequency {\em for charge exchange alone}. The reason for this will become clear soon. So we use the usual formula $\nu_{i, ex}=\sigma_{ex} n_n \vti$. Here, $n_n$ is the number density of neutral atoms,  $\vti^2=\kappa T_i/m_i$, and for $\sigma_{ex}$ we use data from the line 2 in Fig.~\ref{f2}. For plasma parameters in the lower solar atmosphere we use the well known data  for the quiet solar atmosphere \citep{fon}.

\begin{figure}
   \centering
   \includegraphics[height=6.5cm,bb=15 14 272 224,clip=]{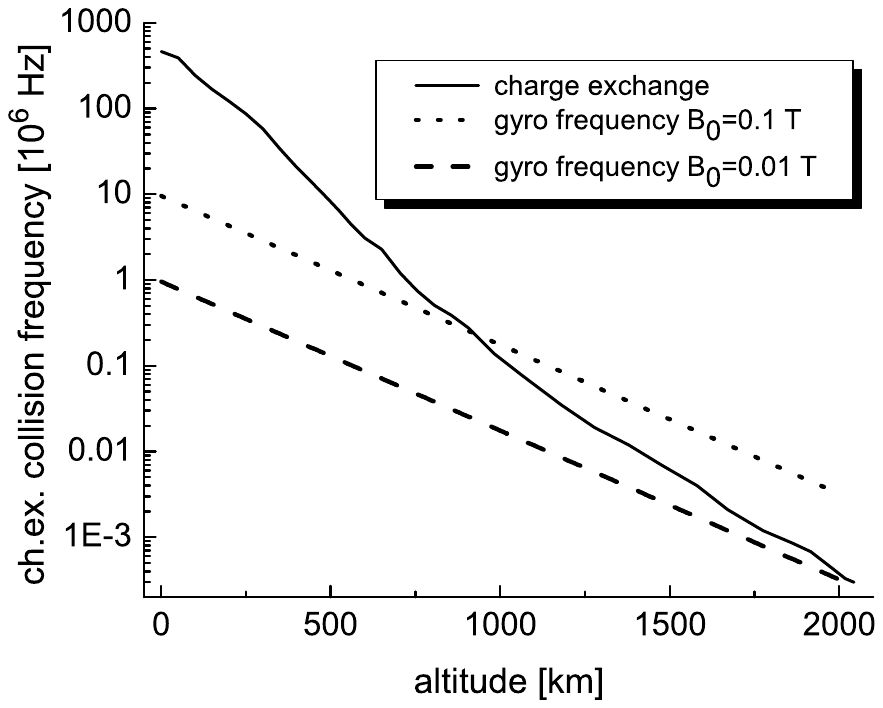}
   \caption{Proton collision frequency for charge exchange $\nu_{i, ex}$ in the lower solar atmosphere in terms of the altitude. Straight dot-line and dash-line are gyro frequency for $B=0.1, 0.01$ T, respectively. }\label{f3}
       \end{figure}

 The result is presented in Fig.~\ref{f3}. Note that we are in a very narrow low-energy range with those quantum fluctuations, and this is why the full line has those small bumps on it. The graph shows that, for an average proton, the charge exchange collision frequency changes from $4.6\cdot 10^8$ Hz at the altitude $h=0$ km, to around 300 Hz at $h=2000$ km. For comparison we give the proton gyro frequency $\Omega_i=eB/m_i$ for $B=B_0\exp[-h/250]$, and for the two starting values of the field $B_0=0.1, 0.01$ T. They are presented by the dashed and dotted lines.

 It is seen that protons are unmagnetized in a large part of the lower solar atmosphere. We stress  that the collision frequency for total elastic scattering (which necessarily includes the charge transfer as well) is in fact higher by about a  factor 4. So the magnetization is in fact even weaker than in Fig.~\ref{f3}, and it extends to even higher altitudes  \citep{v1}.

These numbers have a profound physical meaning. They suggest that a proton in the photosphere changes its identity more than hundred million times per second!  The nature of the process implies that Hydrogen atoms also change identity and this  happens with the frequency $\nu_{n, ex}=\nu_{i, ex}n_i/n_n$. For neutrals, the charge-exchange frequency  \citep{v1} at several altitude is presented in Table~\ref{t1}.

\begin{table}

  \caption{Charge exchange collision frequency for neutral hydrogen at several altitudes in the solar atmosphere. }
  \label{t1}
\begin{tabular}{llccccr}
  \hline
$h$ [km] $\!\! $& $0$ $\!\! $& $150$ $\!\! $& $250$ $\!\! $ & $525$ $\!\! $& $ 1065 $ $\!\! $ & $ 1580 $ $\!\! $  \\
\\
\hline
\\
$\nu_{ni, ex}$ [Hz] $\!\! $ & $2.4\cdot 10^5$ $\!\! $& $ 1073 $ $\!\! $ & $ 184 $ $\!\! $ & $ 8 $ $\!\! $ & $ 376 $ $\!\! $ & 265 $\!\! $  \\
   \hline
\end{tabular}
\end{table}

It is obvious that for both protons and Hydrogen atoms the typical charge exchange time $\tau_{ex}$ is usually many orders of magnitude shorter than any realistic period of Alfv\'{e}n waves which can be expected in such an environment,
\be
\tau_{ex}\ll \tau_{\sss A}. \label{con}
\ee
In view of the results presented in Fig.~\ref{f3} and in Table~\ref{t1}, we come to the following:

 {\em Consequence 3:}  Both ions and atoms in the lower solar atmosphere change identities so frequently that it is impossible to say which is which. For any practical purposes, i.e., for time scales far exceeding this identity-change time scale $1/\nu_{ex}$ (the largest being $1/8$ s for neutrals at the temperature minimum), it is completely obvious that, in average, all particles will respond to electromagnetic perturbations in nearly the same manner.
  The nature of all particles is intermittent; they spend one part of the time in charged state and one part in neutral state.

 {\em Consequence 4:} This further implies that for processes with typical time scales far exceeding $1/\nu_{ex}$, the effective ion plasma density must be  $n_{ef}=n_i+ n_n$. This fact must be taken into account, for example, in the estimates of the Alfv\'{e}n wave flux from the lower solar atmosphere.

These issues will be discussed in more details farther in the text.

\section{Effective charge on particles}
To proceed, we  have to find the effective or mean charge on particles, which is directly caused by the charge exchange effect. This can be done as follows.

\begin{figure}
   \centering
    \includegraphics[height=6cm,bb=17 12 284 222,clip=]{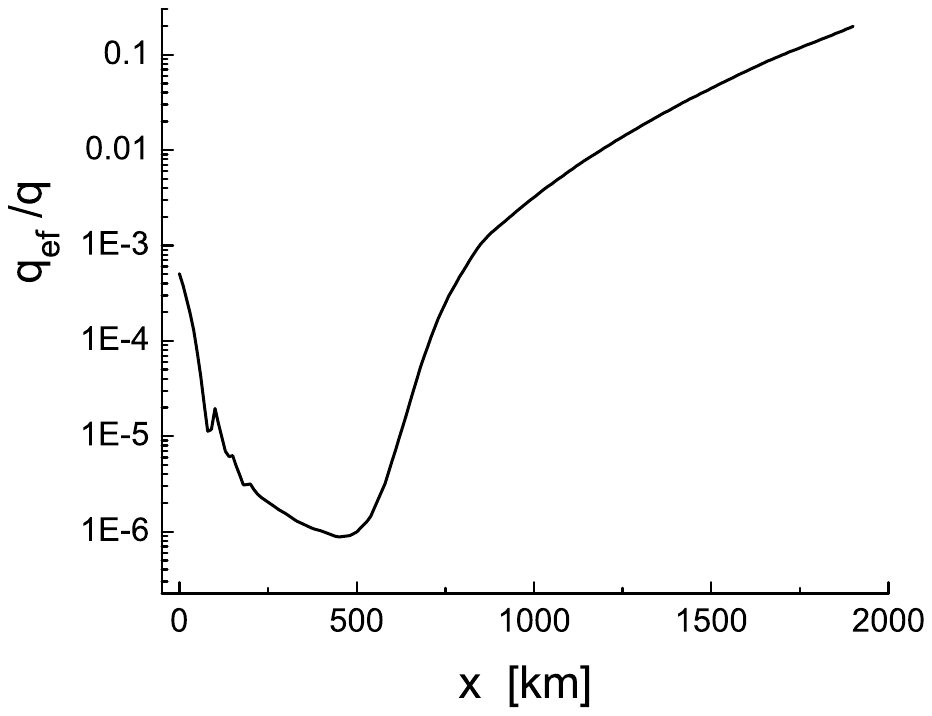}
   \caption{Normalized effective charge on a particle in a proton-Hydrogen mixture, in terms of the altitude.}\label{qef}
       \end{figure}

For each individual particle in the system there is a characteristic time interval $\tau_c=\tau_1+ \tau_2$ within which it passes through two cycles: i) when it is charged ($\tau_1$ interval), and,  ii) when it is in the neutral state ($\tau_2$ interval). After these two cycles the process is repeated,  so we can focus on the $\tau_c$ interval only.  Because of the charge flip-flop, an effective mean charge of such a particle is
\[
q_{ef}=\frac{1}{\tau_c} \int_0^{\tau_c} q f(t), \quad \tau_c=\frac{1}{\nu_{in}} + \frac{1}{\nu_{ni}},
\]
where  we have a step-function $f(t)= 1$ in the starting interval  $\tau_1\in (0, 1/\nu_{in})$, and $f(t)= 0$ in the remaining interval of the cycle $\tau_2\in (1/\nu_{in}, 1/\nu_{ni})$. Note, the collision frequencies here describe the charge-exchange. This yields an effective particle charge in the interval $\tau_{ex}$:
\be
q_{ef}=\frac{q}{1+\nu_{in}/\nu_{ni}}= \frac{q}{1+n_n/n_i}.\label{ef}
\ee
Without neutrals (i.e., in the upper layers) we have $q_{ef}=q$, as it should be. But for a weak ionization the effective charge is very small. This is due to the fact that, in average, a particle spends (much) more time in the neutral state. So  in the photosphere $q_{ef}/q$ is a very small parameter which is spatially dependent, and it  changes rapidly with the altitude. At the temperature minimum ($h=490$ km) its value is $q_{ef}/q\approx 10^{-6}$, and in such an environment slow electromagnetic perturbations are not expected to considerably affect the particle dynamics.  On the other hand, at the altitudes $h=1580, \, 2018$ km, the values of the parameter are, respectively, $q_{ef}/q=0.06, \,0.33$. This normalized effective charge is presented in terms of the altitude in Fig.~\ref{qef}.

{\em Consequence 5:} The effective charge applies to each and every ion and atom in the system. Therefore for any process with the characteristic time $\tau\gg \tau_c$ all particles are subject to electromagnetic forces in the same manner. This implies that  the usual separation of the population into ions (reacting on electromagnetic perturbation) and neutrals (which initially remain static in the presence of such perturbations and later move due to friction by ions) does not hold any longer. The total electromagnetic force acting on {\em all particles} in a unit volume is proportional to $q_{ef} n_{i,ef}$.

The mean motion of all particles is thus the same and in practical sense {\em there is no friction} in the system.  The single fluid of ions and neutrals as a whole can now be (weakly) affected by the magnetic field  as friction between them is absorbed into the one fluid. So, as long as the effects of the charge exchange are not included in a self-consistent manner, every analytical model will fail  in describing  electromagnetic perturbations in such an environment.

\section{The issue of magnetization}
The collisions (of various kinds, both elastic and inelastic) still remain as a fact, and charged particles may be unmagnetized. Following the usual wisdom they should not be  directly affected by the magnetic field,  but see the comment about the Hall effect earlier in the text, Sec.~1. This is the case in the lower solar atmosphere and it may be the same in any other partially ionized plasma. In the photosphere, the total elastic scattering collision frequency may far exceed the gyro-frequency (and this may even happen with electrons), as it is demonstrated in our recent references \citep{v1, v2}.

However, it is claimed that magnetic waves do exist and that they propagate in the photosphere and chromosphere \citep{jes, kos}. If this is a fact then we are definitely still missing a complete picture, and the understanding of magnetic waves  in this weakly ionized environment remains elusive and unexplained.

The purpose of this work is to get some physical insight into the physics of waves in such an environment, and to find mechanisms which might allow for their existence.

\subsection{Rotation of particle displacement plane in the presence of an Alfv\'{e}n wave}
\begin{figure}
   \centering
    \includegraphics[height=9cm,bb=14 14 178 229,clip=]{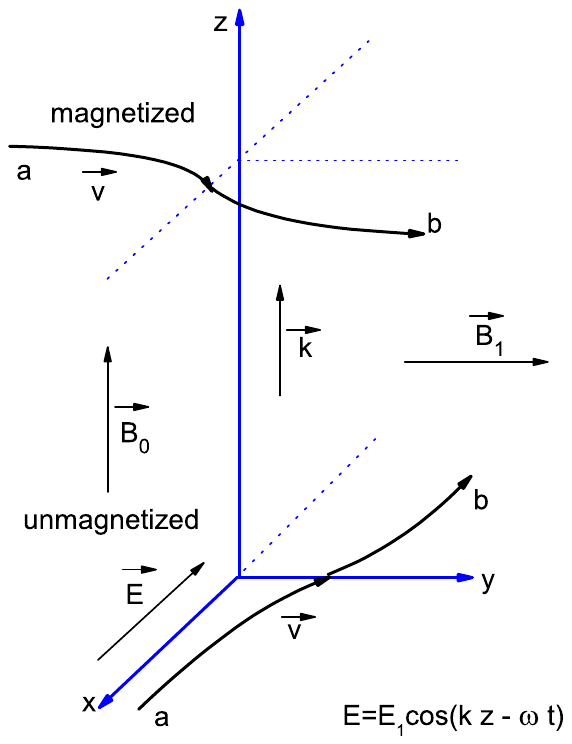}
   \caption{Twist in the motion of the fluid element in the presence of an Alfv\'{e}n  wave which occupies layers of unmagnetized plasma (the lower region of the picture) and  magnetized one (the upper part of the picture). The motion within a half period of the wave is presented. }\label{f4}
       \end{figure}

 In general, we may have two possible cases: with magnetized, and with  unmagnetized particles. Let us assume that the background magnetic field is in $z$-direction, and it is perturbed in $y$-direction. This implies a perturbed electric field in $x$-direction in accordance with Faraday law. This geometry is presented in Fig.~\ref{f4}, and more details are given below in the text.

\paragraph{Unmagnetized particles.} In the case of unmagnetized particles (the collision frequency $\nu_j$ above the gyro-frequency $\Omega_j$) the effect of the Lorentz force is negligible. The particles are never able to perform a gyro-rotation, therefore their motion cannot have a drift character. In other words, the particle predominantly moves {\em in the direction of force}, which is the electric field force in the present case.

But in view of the discussion presented in the text above, due to the charge exchange, this applies to all of them. We are dealing with a mean force acting on the whole fluid. This motion is depicted in Fig.~\ref{f4} by the lower part of the graph.

\paragraph{Magnetized particles} When particles are magnetized, $ \nu_j< \Omega_j$, their dominant motion is due to the $\vec E\times\vec B$-drift \citep{ch}, and this drift motion is the same for both electrons and ions. The motion in the direction of force is only due to polarization drift \citep{v2}. Compared with the leading drift motion, the polarization drift is very small and the ratio of the two displacements is of the order of $\omega/\Omega_j\ll 1$.

So when an Alfv\'{e}n  wave propagates from below through the upper solar layers, {\em there must be a 90 degrees change of the displacement plane of the fluid} as it passes through regions with unmagnetized and magnetized particles. This is depicted in Fig.~\ref{f4}, compare the lower and the upper graphs describing the particle speed.  This rotation of the displacement plane happens only once, when particles change their feature from being un-magnetized to being magnetized. Observers should be able to detect such a rotation.

This motion of {\em magnetized particles} can be described by starting from the momentum equation of the ion fluid, after keeping only terms essential for this discussion. Note that we are now in the frequency range when the collisions are relatively weak and we shall assume that the charge exchange plays no big role, so we have the usual friction force in the ion equation:
\[
m_i\frac{d\vec v_i}{dt}  = e \left(\vec E + \vec v_i\times \vec B\right) - m_i  \nu_{ie} (\vec v_i - \vec v_e)
\]
\be
 - m_i \nu_{in} (\vec v_i - \vec v_n).\label{e1}
 \ee
 Ion-electron friction is omitted here only for simplicity, the mass difference is such that friction with electrons is of a secondary importance. So to explain the essential physics which is in focus here, we keep only the leading order effects. The complete friction force has indeed been studied in our recent works \citet{v2, v3}.

Performing just the vector transformations, from Eq.~(\ref{e1}) the linearized perpendicular speed of ions becomes:
\[
\vec v_{\bot i} = \alpha_i \left( -\frac{1}{B_0} \vec e_z\times  \vec E_1 +
\frac{1}{\Omega_i} \vec e_z\times \frac{\partial \vec v_{\bot i}}{\partial t} + \frac{\nu_{in}}{\Omega_i} \frac{\vec E_1}{B_0}  \right.
\]
\be
\left.- \frac{\nu_{in}}{\Omega_i^2} \frac{\partial \vec v_{\bot i}}{\partial t} +
\frac{i \nu_{in}^2}{\Omega_i^2}\frac{\nu_{ni}}{\omega_n} \vec v_{\bot i} -
\frac{i \nu_{in} \nu_{ni}}{\Omega_i \omega_n} \vec e_z \times \vec v_{\bot i} \right).\label{v3}
\ee
Here, $\omega_n=\omega+ i \nu_{ni}$, and $\alpha_i=1/(1+ \nu_{in}^2/\Omega_i^2)\leq 1$. We have used the momentum equation for neutrals assuming that they are set into motion by friction with ions only \citep{v2}, which gave:
 \be
 \vec v_{\bot n}=\frac{i \nu_{ni}}{\omega+ i \nu_{ni}} \vec v_{\bot i}. \label{neutp}
 \ee
For small ratio $\nu_{in}^2/\Omega_i^2$  (magnetized particles), in average collisions will not significantly affect the particles. Their motion will have drift character, and the dominant term in Eq.~(\ref{v3}) is the first term (the $\vec E\times \vec B$-drift).

Particles move predominantly perpendicular to the electric force, therefore their dynamics is mainly in the $y$-direction. This leading term may be replaced into the remaining terms on the right-hand side of the recurrent formula (\ref{v3}), and they become determined through the electric force. Note, the second term is the polarization drift, which is in the direction of the electric field, but this is small in view of the small parameter $(\partial/\partial t)/\Omega_i \approx |\omega|/\Omega_i$. For negligible collisions we would have only these two drifts:
\[
\vec v_{\bot i}= v_{\sss E}\vec e_y+  v_p\vec e_x.
\]
The motion is essentially two-dimensional although the orbits are strongly extended ellipses.

When particles are {\em unmagnetized}, writing the perpendicular speed in the shape (\ref{v3}) will make no sense because Eq.~(\ref{v3}) is a recurrent formula only when $v_{\sss E}\vec e_y$ is the dominant drift. The motion cannot have drift character in the case of unmagnetized particles, so in this case we are supposed to discuss the starting equation (\ref{e1}).

First, we note that the only active force in Eq.~(\ref{e1}) is due to electric field. For unmagnetized particles in the presence of charge exchange, friction terms will vanish in any case, as we claimed earlier. The remaining Lorentz force only acts on particle or fluid element that is already in motion. In other words, the momentum equation containing only the Lorentz force on the right-hand side would have a trivial solution $\vec v_i=0$.

Now, when particles are not magnetized  their dominant motion will be the direct one, in the direction of the electric field vector. Consequently, only after a fluid element or a particle start moving in the $x$-direction, it will start feeling the effects of the Lorentz force term. For a $\cos(\omega t)$-type dependence of the electric field $\vec E$, the maximum displacement in the $y$-direction is expected when the speed is maximum, that is when $\vec E$ is maximum. After that the particle displacement in the $y$-direction becomes smaller and eventually it moves only in the direction of the field $\vec E$ after half-period of the wave. After this, the electric field direction is changed and the particle (or fluid element) is moving back along the same trajectory. So the motion is again two dimensional as in the usual multicomponent description, but the dominant motion is in the direction of force, and not perpendicular to it as in the case of magnetized species.

To the best of our knowledge, the twist presented in Fig.~\ref{f4} has never been discussed or suggested in the literature. This is a new result which may be applicable even in completely different situations, i.e., in fully ionized plasmas with frequent collisions. The usual wisdom about   Alfv\'{e}n  wave propagation \citep{ch} implies a dominant drift motion of particles, and as such it cannot explain their presence in an environment with unmagnetized species.

\section{Momentum equation in the presence of charge exchange}

In every moment the  charge quasineutrality is satisfied, so in the present case this condition reads
\[
e n_e \equiv e n_i= q_{ef}n_{ef},
\]
where  the effective number of charged particles is $ n_{ef}=n_i+ n_n$.
For low frequency processes, with characteristic time scales $\tau\gg \tau_c$, the correct momentum equation in the presence of charge-exchange and for a plasma with ions and their parental atoms reads:
\be
m_in\frac{d \vec v}{dt}  =  q_{ef}n_{ef}\left(\vec E + \vec v\times \vec B\right)\equiv q n_i\left(\vec E + \vec v\times \vec B\right).\label{e22}
 \ee
Here, $n\equiv n_i+ n_n=n_{ef}$ and $\vec v$ denotes the mean speed of both ions and their parental atoms. We have omitted the {\em friction with electrons} only for simplicity because it is not relevant for the charge exchange discussion. The number of electrons is and their distribution function are  not changed in the process. But this term should be kept in a proper multi-component description.

The equation shows that in an ions-atoms  mixture in any moment the total EM force per unit volume applies only on particles that are in that moment charged. Obviously, in the limit of vanishing neutrals (fully ionized plasma), Eq.~(\ref{e22}) reduces to the usual ion momentum equation.

Note that this equation can be derived from first principles, starting from kinetic equations for a single heavy component (ions plus neutrals) and assuming that all of them have the effective charge $q_{ef}$,  and that the number density $n_{ef}=n$.

Two alternative ways of its  formal derivation are  presented in Secs.~6,~7 below.

\section{ Formal derivation of momentum equation}
In a plasma where charge exchange occurs frequently, ions and neutrals are replacing each other in a very short time interval, so the equations of motion for ions and neutrals may be unified for the $j$th particle as:
\be
 \frac{d{\vec x}_j}{dt} = {\vec  v}_j,
 \label{k1}
 \ee
 \be
 \frac{d{\vec  v}_j}{dt} = \theta(t) \frac{e}{m}\left[ {\vec
E}({\vec  x}_j, t) + {\vec  v}_j \times {\vec  B}({\vec  x}_j, t)\right].\label{k2}
\ee
Here $j=1, 2, \cdots, N_i, N_i+1, \cdots, N_i+N_n$, and $\theta(t)$ is a random
function defined by
 \be
  \theta(t)=\left\{ \begin{array}{l c l} 1 & \mbox{for} & t \in S_i\\ 0 & \mbox{for} & t \in S_n, \end{array}\right.,\label{k3}
\ee
and $S_i$ and $S_n$ are the charged state and neutral state, respectively. The masses of ions and neutrals are assumed to be equal, which is appropriate for a proton-Hydrogen plasma. Note that
the number densities of ions and neutrals are preserved and the total charge density is as well.
 The electromagnetic fields are determined self-consistently by the Maxwell equations.

A microscopic density distribution function for the unified ion-neutral particles is
introduced by
\be
F({\vec  x}, {\vec  v}, t) = \frac{1}{V} \sum_{j=1}^{N_i+N_n} \delta({\vec  x} - {\vec  x}_{ j}(t))\delta({\vec  v} - {\vec  v}_{ j}(t)),\label{k10}
\ee
which allows us to express the current and density. 
We can further write
\[
\frac{\partial}{\partial t}F({\vec  x}, {\vec  v}, t) = \frac{1}{V} \sum_{i=1}^{N_i+N_n}\frac{\partial}{\partial t}\delta({\vec  x} - {\vec  x}_i(t))\delta({\vec  v} - {\vec  v}_i(t))
\]
\[
= \frac{1}{V} \sum_{i=1}^{N_i+N_n}\left( -\frac{d{\vec  x}_i}{dt}\cdot \frac{\partial}{\partial {\vec  x}} - \frac{d{\vec  v}_i}{dt}\cdot \frac{\partial}{\partial {\vec  v}}\right)\delta({\vec  x} - {\vec  x}_i(t))\delta({\vec  v} - {\vec  v}_i(t))
\]
\[
= \left[ -{\vec  v}\!\cdot\! \frac{\partial}{\partial {\vec  x}} - \theta(t) \frac{e}{m}\left({\vec  E} + {\vec  v} \times {\vec  B}\right)\!\cdot\! \frac{\partial}{\partial {\vec  v}}\right] F_{\alpha}({\vec  x}, {\vec  v}, t).
\]
So this yields
\be
 \frac{\partial F_{\alpha}}{\partial t}+{\vec  v}\cdot
\frac{\partial F_{\alpha}}{\partial {\vec  x}} + \theta(t) \frac{e}{m}\left({\vec  E} + {\vec  v} \times {\vec  B}\right)\cdot \frac{\partial F_{\alpha}}{\partial {\vec  v}} =0. \label{k13}
\ee
Since the density distribution function $F$ is not continuous and depends on the initial conditions of particles $\{ {\vec x}_{ j}(0), {\vec v}_{ j}(0)\}$, we introduce an ensemble averaged distribution function over the most probable distribution of the initial conditions as
\be
f({\vec x}, {\vec v}, t) = < F({\vec x}, {\vec v}, t)>,\label{k14}
\ee
where $<F>$ denotes an ensemble average over the initial conditions.
Then the particle discreteness is expressed by
\be
\delta F({\vec x}, {\vec v}, t) = F({\vec x}, {\vec v}, t) - <F({\vec x}, {\vec v}, t)>.\label{k15}
\ee
Obviously
\[
 <\delta F({\vec x}, {\vec v}, t) >=0.
  \]
Since ${\vec E}$ and ${\vec B}$ depend on electrons as well, we can also introduce macroscopic and microscopic fields as
\[
{\vec E}=<{\vec E}> + \delta {\vec E}, \quad {\vec B}=<{\vec B}> + \delta {\vec B}.
\]
To simplify notation, in the following we use ${\vec E}$ and ${\vec B}$ instead of $<{\vec E}>$ and $<{\vec B}>$.
Taking the ensemble average we obtain:
\be
 \frac{\partial f}{\partial t}+{\vec  v}\cdot \frac{\partial f}{\partial {\vec  x}} + \theta(t) \frac{e}{m}\left({\vec  E} + {\vec  v} \times {\vec  B}\right)\cdot
\frac{\partial f}{\partial {\vec  v}} =\left(\frac{\partial f}{\partial t}\right)_c,\label{k16}
 \ee
 where
\be
\left(\frac{\partial f}{\partial t}\right)_c=-<\theta(t) \frac{e}{m}\left(\delta{\vec  E} + {\vec  v}
\times \delta{\vec  B}\right)\cdot \frac{\partial}{\partial {\vec  v}} \delta F>.\label{k17}
\ee
The right hand side in Eq.~(\ref{k16}) represents the interaction among discrete particles through the electromagnetic fields, that is, collisional effects due to the discreteness. When the right hand side is omitted, the remaining equation is a generalized Vlasov equation for plasmas with frequent charge exchange phenomena.

Introducing a mean density and a mean velocity as
\[
 n = \int f d{\vec  v} = n_i + n_n, \quad n{\vec  v} = \int {\vec  v} f d{\vec  v},
 \]
 and  taking the first and second moments of the kinetic equation (without the collision terms for simplicity) gives
\be
\frac{\partial n}{\partial t} + { \nabla}\cdot(n{\vec  v}) = 0,
\label{k18}
\ee
\be
\frac{\partial}{\partial t}(n{\vec  v}) + {  \nabla}\cdot(n{\vec  v}:{\vec
v}) = \theta(t)\frac{ne}{m}\left({\vec  E} + {\vec  v}\times {\vec  B}\right).
\label{k19}
\ee
 When the two obtained equations are combined we obtain:
 \be
  \frac{\partial {\vec  v}}{\partial t} +
{\vec  v}\cdot{  \nabla} {\vec  v} = \theta(t)\frac{e}{m}\left({\vec  E} + {\vec  v} \times {\vec  B}\right). \label{k19}
\ee
We may take an average over the time longer than the characteristic time of charge exchange,  and in this case
\[
 {\bar \theta}(t) = \frac{n_i}{n_i+ n_n}.
\]
Eq.~(\ref{k19}) is the same as Eq.~(\ref{e22}). This is a completely new result applicable for all processes satisfying the condition (\ref{con}).
Yet another way of deriving the common momentum equation for ions and their parental atoms is presented in the next section. 

\section[]{ An alternative derivation of the common momentum equation (6)  }

We shall show that the momentum equation (\ref{e22}), which describes the common dynamics of all heavy particles (ions and neutrals) in the presence of frequent charge exchange,  can be formally derived even by starting from the two separate momentum equations for ions and neutrals.

\be
 \frac{\partial f_i}{\partial t} + {\vec v}\cdot {\nabla} f_i +\frac{e_i}{m_i}\left[{\vec E}({\vec x},t) + {\vec v} \times {\vec B}({\vec x},t)\right]\cdot \frac{\partial f_i}{\partial {\vec v}}=\left(\frac{\partial f_i}{\partial t}\right)_c,\label{a1}
 \ee
 \be
\frac{\partial f_n}{\partial t} + {\vec v}\cdot { \nabla} f_n =\left(\frac{\partial f_n}{\partial t}\right)_c,\label{a2}
\ee
where $f_{\alpha}({\vec x}, {\vec v}, t)$ is an ensemble averaged distribition function of $\alpha$ species,
\be
f_{\alpha}({\vec x}, {\vec v}, t)= \frac{1}{V}<\sum_{j=1}^{N_{\alpha}} \delta({\vec x}-{\vec x}_j(t))\delta({\vec v}-{\vec v}_j(t))>,\label{a3}
\ee
and $(\partial f_{\alpha}/\partial t)_c$ is a collision integral. In general electrons are involved in the collision integral which may be neglected for the present case and  for charge exchange  the density and momentum are conserved, that is,
\[
\int \left(\frac{\partial f_{\alpha}}{\partial t}\right)_c d{\vec v}=0, \quad \int {\vec v}\left(\frac{\partial f_{\alpha}}{\partial t}\right)_c d{\vec v}=0.
\]
Here we introduce a two-body distribution function by
\[
f_{in}\left({\vec x}, {\vec x}', {\vec v} , {\vec v}', t\right)=f_i\left({\vec x}, {\vec v}, t\right)f_n\left({\vec x}', {\vec v}', t\right),
\]
and we get
\[
\left\{\!\frac{\partial }{\partial t} +{\vec v}\!\cdot\! { \nabla} + {\vec v}'\!\cdot\!{ \nabla}' + \frac{e_i}{m_i}\!\left[\!{\vec E}({\vec x}, t) + {\vec v}\!\times\! {\vec B}\left({\vec x},t\right)\!\right]\!\cdot\!\frac{\partial}{\partial {\vec v}}\!\right\}\!f_{in}
\]
\be
=\left(\frac{\partial f_{in}}{\partial t}\right)_c\!, \quad \!\left(\!\frac{\partial f_{in}}{\partial t}\!\right)_c\! =\! f_n \left(\frac{\partial f_i}{\partial t}\right)_c  + f_i \left(\frac{\partial f_n}{\partial t}\!\right)_c\!,\label{ax}
\ee
where
\be
\quad \int f_{in}\left({\vec x}, {\vec x}', {\vec v}, {\vec v}', t\right)d{\vec x}d{\vec x}'d{\vec v}d{\vec v}' = n_{i}n_n.\label{a8}
\ee
For a two-species system with the same number of particles the two-body distribution function is expressed due to the one-to-one correspondence by
\[
f({\vec x}, {\vec x}', {\vec v}, {\vec v}', t )=nf({\vec x}, {\vec v}, t)<\delta({\vec x}-{\vec x}') \delta({\vec v}-{\vec v}')>
\]
\be
+C({\vec x}, {\vec x}', {\vec v}, {\vec v}', t),\label{a9}
\ee
where the second term of RHS is a correlation function. However, in the case studied here, in general
 the numbers of ions and neutrals are different and Eq.~(\ref{a9}) is not straightforwardly applicable to the present case.
So now we introduce a center of density coordinate system through
\[
{\vec R}=\frac{n_i{\vec x}+ n_n{\vec x}'}{n_i+n_n}, \quad {\vec U}=\frac{n_i{\vec v}+n_n{\vec v}'}{n_i+n_n},
 \]
 \[
 {\vec r}=\frac{2\left(n_i{\vec x} -n_n {\vec x}'\right)}{n_i+n_n} , \quad  {\vec u}=\frac{2\left(n_i{\vec v} - n_n {\vec v}'\right)}{n_i+n_n},
\]
This transforms Eqs.~(\ref{ax}, \ref{a9}) into
\[
\bigg\{\frac{\partial }{\partial t} + {\vec U}\cdot \frac{\partial}{\partial{\vec R}} + {\vec u}\cdot \frac{\partial}{\partial{\vec r}}
\]
\[
 + \frac{e_i}{m_i}\frac{n_i}{n_i+n_n}\bigg[{\vec E}\left(\frac{n_i+n_n}{2n_i}\left({\vec R}+\frac{{\vec r}}{2}\right), t\right)
 \]
 \[
 + \frac{n_i+n_n}{2n_i}\left({\vec U}+\frac{{\vec u}}{2}\right)\!\times\! {\vec B}\left(\frac{n_i+n_n}{2n_i}\left({\vec R}+\frac{{\vec r}}{2}\right)\!,t\right)\!\bigg]
 \]
 \[
 \cdot \left(\frac{\partial}{\partial {\vec U}} + 2\frac{\partial}{\partial {\vec u}}\right)\bigg\}f_{in}=\left(\frac{\partial f_{in}}{\partial t}\right)_c,
 \]
\[
 f_{in}d{\vec x}d{\vec v}d{\vec x}'d{\vec v}'
 = \left[ \frac{n_in_n}{n_i+n_n} F_{in}\left({\vec R}, {\vec U}, t\right)\right.
 \]
 \[
 \left.
 \times
  <\delta\left({\vec r} -2\frac{n_i-n_n}{n_i+n_n}{\vec R}\right)\delta\left({\vec u}-2\frac{n_i-n_n}{n_i+n_n}{\vec U}\right)>\right.
  \]
  \be
  \left.
  + C_{in}\left({\vec R}, {\vec U}, {\vec r}, {\vec u}, t\right)\right]d{\vec R}d{\vec U}d{\vec r}d{\vec u},\label{a12}
\ee
where we have used
\[
{\vec x} - {\vec x}' = \frac{(n_i+n_n)^2}{4n_in_n}\left({\vec r} - 2\frac{n_i-n_n}{n_i + n_n}{\vec R}\right).
 \]
The volume element of the phase space is preserved as
\[
f_{in}({\vec x}, {\vec x}',{\vec v}, {\vec v}',  t)d{\vec x}{\vec x}'d{\vec v}d{\vec v}'\!=\! F_{\alpha}({\vec R}, {\vec U}, {\vec r}, {\vec u}, t)d{\vec R}d{\vec U}d{\vec r}d{\vec u}.
\]
The coefficient of $F_{in}({\vec R}, {\vec U}, t)$ in the RHS of Eq.~(\ref{a12}) is uniquely determined by the condition that it is symmetric with respect to the species and  proportional to $N$.  From Eqs.~(\ref{a8}, \ref{a12}) we have
\begin{equation}
\int F_{in} d{\vec R} d{\vec U}=n_i + n_n.
\end{equation}

When charge exchange is so frequent that ions and neutrals are supposed to behave together, the second term of the RHS of Eq.~(\ref{a12}) is neglected compared with the first term. Integrating Eq.~(\ref{a12}) over ${\vec r}$ and ${\vec u}$ yields
\be
\left[\frac{\partial }{\partial t} + {\vec U}\cdot \frac{\partial}{\partial{\vec R}} + \frac{\vec {\mathcal F}}{m_i}\cdot \frac{\partial}{\partial {\vec U}} \right]F_{in}({\vec R}, {\vec U}, t)=0,\label{af}
\end{equation}
\[
\vec {\mathcal F}=e_i\frac{n_i}{n_i+n_n}\bigg[{\vec E}({\vec R}, t)
+ {\vec U} \times {\vec B}({\vec R}),t)\bigg],
\]
where we have used
\begin{equation}
 \int \left(\frac{\partial f_{in}}{\partial t}\right)_c d{\vec u} d{\vec r}=0.
 \end{equation}
Eq.~(\ref{af}) is a generalized Vlasov kinetic equation for the species with an effective charge $q_{ef}$ which we introduced earlier in the text, see Eq.~(\ref{ef}). So now, following the well-known procedure, it directly yields all fluid equations as zero-momentum (continuity equation), the first momentum (equation of motion (\ref{e22})), etc.

\section{Application to Alfv\'{e}n waves}

In Fig.~\ref{f4}, a plain-polarized wave is assumed just to explain the basic features, which is fine most of the time for large wavelengths $k\lambda_i\ll 1$, $\lambda_i=c/\omega_{pi}$, $\omega_{pi}^2=e^2 n_{i0}/(\varepsilon_0 m_i)$. But in general any direction of the perturbed perpendicular field is possible (i.e., we should allow for a circular polarization). So we may have  $\vec B_1=B_x\vec e_x+  B_y\vec e_y$ and similar for the electric field.

Using momentum equation for inertialess electrons we have
\be
\vec E=-\vec v_e\times \vec B.\label{d1}
\ee
In the present case, for the time scales determined by (\ref{con}), the current is produced by all heavy species and by electrons, $\vec j=q_{ef} n_{ef} \vec v - e n_e \vec v_e$, which with the help of Amp\`{e}re law yields
\be
\vec v_e=\vec v - \frac{1}{\mu_0 e n_i}\nabla\times \vec B. \label{d2}
\ee
This is used in Eq.~(\ref{d1}) and the resulting expression is further used in the Faraday law which finally yields
 \be
\frac{\partial \vec B}{\partial t}=\nabla \times \left(\vec v\times \vec B\right)
-\frac{1}{\mu_0e n_i}\nabla \times \left[\left(\nabla \times \vec B\right)\times \vec B\right]. \label{d3}
\ee
We now linearize  our new momentum equation (\ref{e22}), and we eliminate the electric field with the help of  the Faraday law $E_x=\omega B_y/k$, $E_y=-\omega B_x/k$,
\be
\vec v=\frac{i e n_i}{m_i \omega n} \left[\left(\frac{\omega B_y}{k} + v_y B_0\right)\vec e_x - \left(\frac{\omega B_x}{k} - v_x B_0\right)\vec e_y\right].
\label{d4}
\ee
We further linearize Eq.~(\ref{d3}), and we have a closed set for $\vec B_1, \vec v_1$. This yields the following dispersion equation:
\be
\omega^4 - 2 \omega^2 k^2 v_a^2 \left[1+ \left(1+ \frac{n_{n0}}{n_{i0}}\right) \frac{k^2\lambda_i^2}{2}\right] + k^4 v_a^4=0. \label{d5}
\ee
Here,
\be
v_a^2=B_0^2/[\mu_0 m_i(n_{i0}+  n_{n0})] \label{as}
\ee
 is the Alfv\'{e}n speed which includes both ions and neutrals.
This can be written as:
\[
\omega^2= k^2 v_a^2 \left\{1 + \left(1+\frac{n_{n0}}{n_{i0}}\right) \frac{k^2 \lambda_i^2}{2} \pm \left(1+\frac{n_{n0}}{n_{i0}}\right)^{1/2} k\lambda_i
 \right.
 \]
 \be
 \left.
 \times
 \left[1+ \left(1+\frac{n_{n0}}{n_{i0}}\right)  \frac{k^2 \lambda_i^2}{4}\right]^{\!1/2}\right\}. \label{e8}
\ee
Setting $n_{n0}=0$ yields a solution well-known from the literature \cite{al}, with corrections containing the small parameter $k \lambda_i$ which gives so called left and right circularly polarized waves.  In the usual AW limit  $k\lambda_i\ll 1$ this yields
\be
\omega^2= k^2 v_a^2. \label{e10}
\ee
 In any particular moment, all particles (both charged and uncharged) take part in low-frequency perturbations, and this is reflected in the expression for the AW speed.

\section{\label{p} Proving the theory}

\noindent I. There are numerous reports about the presence of Alfv\'{e}n waves in the lower solar atmosphere. If this is taken as a fact then this  might be understood as a plausibility of the theory presented here because the friction, when calculated in the usual way, destroys them instantly.

\noindent II. The presence of the waves can only be explained with the help of charge exchange, as described with our Fig.~\ref{f4}, which is important for the theory presented here. If such a rotation of the Alfv\'{e}n wave plane could be verified by observations, this alone would mean the verification of the theory as well.

To help observers in detecting the twist presented in Fig.~\ref{f4}, here we give some more details about possible locations (altitudes) where such a transformation can take place. So the wave features depend on the ratio
\be
\nu_i/\Omega_i. \label{rat}
\ee
 Here, $\nu_i$ is the most dominant collision frequency for ions. As shown in \citet{v1}, this can be either $\nu_{in}$ or $\nu_{ii}$, and this is dependent on the altitude. Now, important to stress is that in both cases the collision frequency contains the cross section for total elastic scattering. For ion-neutral collisions this cross section is determined self-consistently and it contains effects of the charge exchange. From Fig.~1 in \citet{v1} it may be seen that elastic scattering cross section is roughly by a factor 4 greater than the charge exchange cross section (which we used in Fig.~\ref{f3} in the present work).

So using very accurate values from  \citet{v1}, and for the magnetic field of the same shape as used previously in the text, $B=B_0\exp[-h/250]$, we can now check at which altitudes the ratio (\ref{rat}) is equal to unity. The wave is expected to change its polarization plane around this altitude. From the graphs in \citet{v1} it may be seen that even assuming an exceptionally strong magnetic field $B_0=0.1$ T, we have $\nu_{in}>\Omega_i$ up to roughly $h=1400$ km. But from this altitude and above, Coulomb collisions become dominant so  $\nu_{ii}>\Omega_i$ and protons remain unmagnetized.

A big drop in the Coulomb collision frequency takes place only somewhere above $h=2400$ km. So only at these altitudes, or higher,  the mentioned twist of the wave plane should be expected.

This all holds on condition that the assumed variation of the magnetic field is valid. But even if we keep the magnetic field $B_0=0.1$ T constant with altitude, the protons will remain unmagnetized for some 800 kilometers. Therefore the change of the wave plane cannot possibly happen below this altitude. We believe these should be  clear indicators for observers; the wave twist should be taking place in the range 800-2400 km in the lower atmosphere. With some more accurate values for the magnetic field the altitude can be pinpointed far more precisely.

\noindent III. The estimated Alfv\'{e}n wave speed which can be found in some references  \citep{jes} is around 22 km/s, with the magnetic field which they assumed to be 0.1 T. Such a speed can only be obtained if the number density involved [see Eq.~(\ref{as}) in the previous section] has the value $n=10^{22}$ m$^{-3}$. We note that the ion number density in the photosphere is several orders of magnitude below this value \citep{fon}, so the density which enters the expression for the Alfv\'{e}n speed is in fact the density of neutrals $n_{n0}$, as expected from our analysis. This looks like a rather convincing proof of our theory.

However, it is necessary to stress that such an expression for the Alfv\'{e}n speed with the neutral density has been used in the past in the usual MHD descriptions as well. In fact, it can even be obtained analytically under certain conditions \citep{kp}. In such a modeling there are two propagation windows for  Alfv\'{e}n waves, determined either by the amount of collisions or equivalently by the wave length. In other words, there is a range of parameters for which the wave cannot propagate and this has been known for more than half a century \citep{kp}. We have discussed these issues in detail in our recent work \citet{v2}. But MHD theory cannot provide a self-consistent  explanation for the issue of magnetization of particles and the presence of waves; it operates with guiding center of mass and not with actual particles. Even drift motion cannot be described within such a rough model. The magnetized particles are an underlying assumption in MHD theory; in other words, plasma is by definition affected by the magnetic field.

Contrary to this, our theory includes the charge exchange, it is based on the single condition (\ref{con}), and the theory works regardless if particles are magnetized or not, hence the predicted twist of the wave plane in Fig.~\ref{f4}. Note also that there are no two propagation windows in our theory, the wave propagates for any parameters.

\noindent IV. Yet another practical way of proving our theory may be deduced by analyzing displacement speed of the plasma involved in wave motion. As discussed in Sec.~4, the nature of displacement depends on magnetization.

In the upper layers with magnetized particles the leading order speed is due the $\vec E\times \vec B$-drift
\be
 v_{\sss E}=\frac{E_1}{B_0}=\frac{\omega}{k}\frac{B_1}{B_0}. \label{p1}
\ee
In this expression it might be  appropriate to set $\omega/k=c_a$, where $c_a=B_0/(\mu_0 m_i n_{i0})$. This because the charged species are subject to drift motion, they are magnetized, which implies that the number of neutrals is considerably reduced so the Alfv\'{e}n speed contains ion density only.

In the lower layers our new momentum equation (\ref{e22}) applies. The dominant motion must be due to the electric field force. The speed of this direct motion from Eq.~(\ref{e22}) is:
\be
v_{dir}=\frac{e n_i}{m_i \omega n} E_1=\frac{\Omega_i}{\omega}\frac{n_{i0}}{n_{i0}+ n_{n0}} \frac{\omega}{k}\frac{B_1}{B_0}.\label{p2}
\ee
In this case we know that  $\omega/k=v_a$, where $v_a$ is given by Eq.~(\ref{as}). Assume now that the wave occupies the whole region, with magnetized particles in the upper layers and with the environment where the charge exchange plays the main role in the lower layers, and that the frequency is the same in both regions. If we further assume that the magnetic field perturbation is the same (which may be a very rough assumption), this yields:
\be
\frac{v_{dir}}{ v_{\sss E}}=\frac{\Omega_i}{\omega}\left(\frac{n_{i0}}{n_{i0} + n_{n0}}\right)^{3/2}. \label{p3}
\ee
Obviously, both terms on the right-hand side are altitude dependent. The minimum value of the second term is $10^{-9}$ but it increases with altitude. The first term can be huge, so the ratio of the two speeds can have any value. For example, taking parameters from \citet{jes}, the speed ratio at the temperature minimum  may become of the order on unity.  But the estimate formula (\ref{p3}) is rough due to various reasons; one obvious is that the ratio $B_1/B_0$ cannot be the same in both regions. The equilibrium magnetic field changes with the altitude and the value for $B_1$ is hard to guess.

 But it is definitely possible to find differences in the displacement speeds for the two regions by using the expressions (\ref{p1}, \ref{p2}) as they are, without further assumptions, and then calculate the speed by using the parameters from observations.

\noindent V. The Alfv\'{e}n speed introduced in Eq.~(\ref{as}) includes the neutral density, which changes for about 6 orders of magnitude. Being under the square root, this implies that the speed changes by a factor 1000 if the magnetic field is constant. Though the magnetic field changes as well, but there exist models for its variation, so the resulting changes in the speed could be measured.

\section{Summary}
The analysis presented here leads to the following conclusions:
\begin{description}
\item{a)} For low-frequency processes, particles (ions and neutrals) in plasma where charge exchange is taking place should be treated as a single fluid with an effective charge  as calculated in the text.
     \item{b)} In a plasma like the solar photosphere, the low-frequency magnetic waves can propagate even if particles are not magnetized. This is due to the fact that it is not possible to separate particles into two different populations as charged and neutral species, therefore there can be no friction force between ions and neutrals in the usual sense of its meaning.

    \item{c)} Because of the charge exchange, all particles take part in low-frequency electromagnetic perturbations. This is the reason why electrodynamic theory is needed in an environment like the photosphere, and not a pure fluid theory as in the terrestrial seas, in spite of the fact that the conductivity in these two environments is practically the same.

The frequent charge exchange induces one fluid of ions and neutrals and as such, collisions between ions and neutrals are absorbed in such a collisionless one fluid. So  the characteristic frequency which needs to be compared with the cyclotron frequency is
no longer the collision frequency  but the fluid oscillation frequency. In this sense one fluid is magnetized (or magnetic) although the charge
density is reduced. Eq.~\ref{e22} is no longer an equation for single particle.

    \item{d)} The effects described here are principally different from the classic MHD interpretation which predicts two propagation windows for the Alfv\'{e}n waves in a weakly ionized environment. Such a behavior within MHD description is caused by the friction between ions and neutrals. The charge exchange effects discussed here have a completely different nature, and the analysis presented in the work suggests, in average, the absence of ion-neutral friction. We expect that the flux of  Alfv\'{e}n waves eventually coming from the photosphere could be considerably modified, and their role in the heating of upper layers should be reconsidered.
    \item{e)} The theory presented here explains why magnetic waves can still propagate in an environment in which ions are obviously not magnetized. In application to the lower solar atmosphere, the results presented here predict that the plane of displacement of the fluid must change by 90 degrees when an Alfv\'{e}n wave propagates from the area where particles are un-magnetized (photosphere) to the area where they are magnetized (chromosphere). Providing some accurate values for the magnetic field, in view of the most accurate cross sections we are using here, and the corresponding collision frequencies, we are able to very accurately determine altitudes at which particle magnetization changes.

\end{description}

In the present study we have been focused on charge exchange as the most dominant of all inelastic collisions, which is in the same time a rather specific inelastic effect because it preserves the number density of all particles involved in it. In application to the lower solar atmosphere, other inelastic collisions, like radiative recombination and 3-body recombination, are known to be far less frequent \citep{vpla}, with maximum frequencies of the order of a few tens of Hz only.

\section*{Acknowledgments}

J.V. acknowledges a) the financial support from the Spanish
Ministry of Economy and Competitiveness (MINECO)
under the 2011 Severo Ochoa Program MINECO SEV-
2011-0187, and b) fruitful discussions with M. Collados and E. Khomenko.

M. Luna acknowledges the support by the Spanish Ministry of Economy and Competitiveness through projects AYA2011-24808, AYA2010-18029 and AYA2014-55078-P. This work contributes to the deliverables identified in FP7 European Research Council grant agreement 277829, ``Magnetic Connectivity through the Solar Partially Ionized Atmosphere'' (PI: E. Khomenko).

\label{lastpage}
\bsp

\end{document}